\documentclass[]{JHEP3}

\usepackage{latexsym} 
\usepackage{amsfonts} 

\usepackage{indentfirst}

\def\d{\delta}

\def\k{\kappa}
\def\la{\lambda}
\def\La{\Lambda}
\def\om{\omega}

\def\t{\tau}
\def\th{\theta}

\def\cN{\mbox{${\cal N}$}}

\def\diag{{\rm diag}}

\def\tr{{\rm tr}}

\def\llap#1{\hbox to 0pt{\hss#1}}

\def\pola{a\llap{\hbox{\char'30\kern-1.2pt}}}
\def\pole{e\llap{\hbox{\char'30\kern-.8pt}}}

\newcommand{\non}{\nonumber\\}

\newcommand{\lb}{\left(}
\newcommand{\rb}{\right)}

\newcommand{\beq}{\begin{equation}}
\newcommand{\eeq}{\end{equation}}
\newcommand{\beqa}{\begin{eqnarray}}
\newcommand{\eeqa}{\end{eqnarray}}
\newcommand{\barr}{\begin{array}}
\newcommand{\earr}{\end{array}}
\newcommand{\ben}{\begin{enumerate}}
\newcommand{\een}{\end{enumerate}}
\newcommand{\bit}{\begin{itemize}}
\newcommand{\eit}{\end{itemize}}

\newcommand{\refeq}[1]{(\ref{#1})}

\newcommand{\aut}{automorphism  } 

\newcommand{\auts}{automorphisms  }
\newcommand{\autsn}{automorphisms}
\newcommand{\rep}{representation }
\newcommand{\repn}{representation}
\newcommand{\reps}{representations }
\newcommand{\repsn}{representations}
\newcommand{\irrep}{irreducible representation }

\newcommand{\irreps}{irreducible representations }
\newcommand{\irrepsn}{irreducible representations}

\newcommand{\hw}{highest weight }

\newcommand{\rhs}{the right hand side }

\def\cG{{\cal G}}
\def\cR{{\cal R}}
\def\cU{{\cal U}}
\def\cC{{\cal C}}

\def\bN{{\mathbb{N}}}
\def\bZ{{\mathbb{Z}}}

\def\x{\times}
\def\ox{\otimes}

\def\ggt{\mathfrak{g}}
\def\hggt{\widehat{\ggt}}
\def\qgt{\cU_q (\ggt)}

\def\emb{\hookrightarrow}

\def\1{\mathbb{I}}
\def\2{[2]_q}
\def\M{\mathbf{M}}
\def\Lp{\mathbf{L^{+}}}
\def\Lm{\mathbf{L^{-}}}
\def\Lpm{\mathbf{L^{\pm}}}
\def\K{\mathbf{K}}
\def\Kp{\mathbf{K^{+}}}
\def\Km{\mathbf{K^{-}}}

\def\R{\mathbf{R}}
\def\X{\mathbf{X}}
\def\s{\mathbf{s}}
\def\t{\mathbf{t}}

\def\bD{\mathbf{D}}
\def\C{\mathbf{C}}

\def\diag{\textrm{diag}}
\def\GcG{\cG_L \otimes_\cR \cG_R}

\newcommand{\lhs}{the left hand side }

\newcommand{\gt}[1]{\mathfrak{#1}}

\newcommand{\ovl}[1]{\overline{#1}}

\newcommand{\ket}[1]{\vert #1 \gg}

\newcommand{\ketC}[1]{\vert #1 \gg_C}
\newcommand{\braCo}[1]{{}^{\om_c}_C\ll #1 \vert}
\newcommand{\ketCo}[1]{\vert #1 \gg^{\om_c}_C}

\newcommand{\qA}[1]{\cU_q (\gt{su}_{#1+1})}
\newcommand{\qUo}[1]{\cU{'}_q (\gt{so}_{#1})}
\newcommand{\Uo}[1]{\cU (\gt{so}_{#1})}
\newcommand{\faffA}[1]{P^{\k}_{+}(A_{#1})}

\newcommand{\reA}[1]{\textrm{REA}_q(A_{#1})}
\newcommand{\treA}[1]{\textrm{tREA}_q(A_{#1})}

\newcommand{\nl}{\newline}

\title{Twisted WZW Branes from Twisted REA's\footnote{Work 
supported by Polish State Committee for Scientific Research 
(KBN) under contract 2 P03B 001 25 (2003-2005)}}
\author{Jacek Pawe\l czyk$^a$, Harold Steinacker$^b$ and 
Rafa\l ~R. Suszek$^{a,c,}$\thanks{Marie Curie Fellow} \\
\normalsize{$^a$ \textit{Institute of Theoretical Physics, 
Warsaw University, ul. Ho\.za 69, PL-00-681 Warsaw, Poland}} \\
\normalsize{$^b$ \textit{Sektion Physik der 
Ludwig--Maximilians--Universit\"at M\"unchen, Theresienstr. 37, 
D-80333 M\"unchen, Germany}} \\
\normalsize{$^c$ \textit{King's College London, Strand, 
London WC2R 2LS, UK}} \vspace{+0.5cm}\\
E-mail: \email{Jacek.Pawelczyk@fuw.edu.pl, 
Harold.Steinacker@physik.uni-muenchen.de, 
Rafal-Roman.Suszek@fuw.edu.pl}}

\abstract{Quantum geometry of twisted Wess--Zumino--Witten 
branes is formulated in the framework of twisted Reflection 
Equation Algebras. It is demonstrated how the \rep theory 
of these algebras leads to the correct classification 
of branes. A semiclassical formula for quantised brane 
positions is derived and shown to be consistent with earlier 
string-theoretic analyses.}


\preprint{ }

\begin{document}

\frenchspacing

\newpage
\section{Introduction.}

Branes on group manifolds and quotients thereof have long 
been at the focus of research efforts aimed at understanding 
the deformation of classical geometry and gauge dynamics 
effected by string propagation in background 
fluxes\footnote{For a review, see: \cite{schom-lect}.}. While 
the branes naturally lead to the concept of a curved 
non-commutative space \cite{fuzzy,twfuzzy}, they are still 
amenable to direct investigation using diverse methods such 
as the Lagrangian formalism of the associated WZW models 
\cite{cohom,lagr-wzw} confining the branes to (twisted) 
conjugacy classes, effective field theory formulated in terms 
of the Dirac--Born--Infeld functional \cite{field-theory} 
proving their stability, matrix models \cite{fuzzy,twfuzzy} 
providing a semi-classical picture of the geometry and gauge 
dynamics, renormalisation group techniques 
\cite{fuzzy,twfuzzy,RGK-theory,rgk} capturing brane 
condensation phenomena, K-theory 
\cite{RGK-theory,mms,K-theory-rest} classifying their charges 
and Boundary Conformal Field Theory (BCFT) offering access 
to their microscopic structure via the boundary state 
construction.

In the latter approach, (twisted) branes are identified 
with states in the Hilbert space of the bulk (or closed 
string) theory implementing (twisted) gluing conditions 
for chiral currents of the bulk CFT ($a$ is an index 
of the adjoint \rep of the horizontal Lie algebra 
$\ggt \equiv Lie G$ of the Kac--Moody algebra 
$\hggt_{\k}$, $n$ enumerates Laurent modes and $B$ is 
a boundary state label): 
\beq\label{glue0}
\lb J^a_n \ox \1 + \om(\1 \ox J^a_{-n}) \rb \ket{B}^{\om} 
= 0,
\eeq
where $\om$ is an outer \aut of the current algebra 
$\hggt_{\k}$ (see, e.g., \cite{auts,twcond}). 
Thus branes break the full chiral symmetry algebra 
$\hggt_{\k}^L \times \hggt_{\k}^R$ of the bulk WZW 
to the subalgebra spanned by annihilators 
of $\ket{B}^{\om}$, isomorphic to $\hggt_{\k}$. 

Non-commutative geometry entered the stage thus set 
in \cite{fuzzy} where a matrix model of "fuzzy" physics 
of untwisted branes was explicitly derived in the large 
volume (or, equivalently, large level $\k$) limit. 
The twisted case was then examined at great length 
in \cite{twfuzzy}, along similar lines. The semi-classical 
approach of \cite{fuzzy} was later extended in \cite{qmod} 
where an Ansatz for brane geometry and gauge dynamics 
at arbitrary level was advanced, based on the fundamental 
concept of quantum group symmetry, as suggested 
by the underlying (B)CFT, and the well-known correspondence 
between untwisted affine Lie algebras $\hggt_{\k}$ 
and Drinfel'd--Jimbo quantum algebras $\qgt$ (see, e.g., 
\cite{press}). The latter proposal was shown to successfully 
encode essential (untwisted) brane data such as tensions, 
localisations, the algebra of functions, internal gauge 
excitations and interbrane open string modes. It was also 
generalised in \cite{qorb} to a class of orbifold backgrounds, 
known as simple current orbifolds $SU(N)/\bZ_N$, whereby 
the basic structure of the associated matrix model, 
a so-called Reflection Equation 
Algebra\footnote{Cp \cite{skl,kss,frt}, see also: 
the Appendix.} (REA) $\reA{N}$, was examined extensively. 
The study revealed an attractive geometric picture 
behind the compact algebraic framework of the REA's, which 
was next exploited in an explicit construction of some new 
quantum geometries corresponding to (fractional) orbifold 
branes.    

One particular aspect of the non-classical WZW geometry is 
quantisation of brane locations within $G$. It can be 
derived rather straightforwardly 
from the relative-cohomological constraints on the background 
fluxes of the relevant Lagrangian boundary WZW model 
ascertaining well-definedness of the associated path integral 
\cite{cohom}. We shall explicitly refer to some results 
of the cohomological analysis in what follows.  

In this paper, we discuss an algebraic framework relevant 
to the analysis at arbitrary level of twisted branes 
on $SU(2n+1)$ group manifolds. Accordingly, we specialise our 
exposition to the case $\hggt_{\k} = A_{2n}^{(1)}$ (in which 
$\om_c$ is the standard $\bZ_2$-reflection of the Dynkin 
diagram). The exposition is centred on the CFT-inspired notion 
of twisted quantum group symmetry, as represented by so-called 
twisted Reflection Equation Algebras (tREA) $\treA{2n}$. 
The latter are directly related to quantum algebras 
$\qUo{2n+1}$, with a known representation theory 
\cite{tongk,tworep,CasoN,haklipo,hapos,klimyk,tensor}. 
The $\qUo{2n+1}$ are (coideal) subalgebras\footnote{Structures 
of this kind have long been known to arise naturally 
in the related context of $(1+1)$-dimensional integrable 
models on a half-line, with involutively twisted gluing 
condition for chiral symmetry currents at the boundary, 
cp \cite{mackay}, see also \cite{delius}.} of $\qA{2n}$ - 
a quantum-algebraic counterpart of the classical subalgebra 
structure: $\gt{so}_{2n+1} \emb \gt{su}_{2n+1}$ 
\cite{nuw,nosu}. Using these facts, we provide evidence 
of an intricate relationship between twisted boundary states 
\cite{twcond,quel,twfus} and the \rep theory 
of the $\om_c$-invariant subalgebra 
$\gt{so}_{2n+1} \cong (\gt{su}_{2n+1})^{\om_c}$, and 
subsequently reconcile our result with the structure 
of the \rep theory of $\qUo{2n+1}$ at $q$ a root 
of unity\footnote{As dictated by the CFT.}, embedded in that 
of $\qA{2n}$. We also rederive the quantisation rule 
for twisted brane positions within the WZW group manifold 
of $SU(3)$ (originally obtained from cohomological analysis 
in \cite{stan}), whereby we establish - in direct analogy 
with the untwisted case - a simple geometric meaning 
of the Casimir operators of $\treA{2n}$.   

Let us now give an outline of the present paper. Section 
\ref{sec:tw-Db}. is a warm-up presentation of the classical 
geometry of the twisted branes. Section \ref{sec:elem}. 
discusses chosen features of the twisted Reflection Equation 
Algebras. Section \ref{sec:tgeom}. contains the main results 
of this work: classification of the twisted branes through 
the \rep theory of the tREA's and a semiclassical derivation 
of the quantisation rule for brane positions in the $SU(2n+1)$ 
group manifold. In the appendices attached, we list further 
properties of Reflection Equations and the $\qUo{2n+1}$ 
algebras.

\section{Classical geometry of twisted WZW 
branes.}\label{sec:tw-Db}

At the classical level, stable branes of the WZW model 
in the Lie group target $G$ are described by (twisted) 
conjugacy classes of the form:
\beq\label{conj-classes-om}
\cC^{\om}(t) = \left\{ h t \om(h^{-1}) \ \big\vert \ h \in G 
\right\},
\eeq
with $t$ in the "symmetric" subgroup $T^{\om}$ of the maximal 
torus $T \subset G$, i.e. $t \in T$ with $\om(t) = t$, whence - 
in particular - the conjugacy classes are invariant under $\om$. 
When $G=SU(2n+1)$ and $\om = \om_c$ (the case of interest) we 
may choose complex conjugation $\rho$ as a group-integrated 
representative of $\om$, whereby the above reduces to ($T$ 
denotes transposition) 
\beq
\cC^{\rho}(t) = \left\{ h t h^T \ \big\vert \ h \in SU(2n+1) 
\right\}.\label{conj-classes}
\eeq 
Let $K_t = \{h \in G:\; h t h^T = t \}$ be the stabiliser 
subgroup (in the twisted adjoint \repn) of $t \in T^\om$. 
For $t = \1$, the stabiliser $K_t$ coincides with 
the group $SO(2n+1)$. In the algebraic setup to be developed, 
we shall encounter a quantum deformation of this group (see 
Sec.\ref{sec:elem}.). Clearly, $\cC^{\om}(t)$ can be viewed 
as a homogeneous space\footnote{In this picture, the map: $G/K_t 
\rightarrow \cC^{\om}(t), \quad h K_t \mapsto h t \om(h^{-1})$ is 
manifestly well-defined and bijective. Note that \lhs is a 
one-sided (right) coset of $G$.}:
\beq
\cC^{\om}(t) \cong G/K_t.
\label{coset}
\eeq
The twisted conjugacy classes are invariant under the twisted 
adjoint action of the vector subgroup 
$G \cong G_V \emb G_L \x G_R$ of the group of symmetries 
of the target manifold,
\beq\label{adj-action}
G \cC^{\om}(t) \om (G^{-1}) = \cC^{\om}(t).
\eeq
This is a classical counterpart of the symmetry breaking pattern: 
$\hggt_{\k}^L \x \hggt_{\k}^R \to \hggt_{\k}$ 
mentioned under \refeq{glue0}. In this context, the distinguished 
character of the $\om$-invariant subgroup derives from the fact 
that a given twisted conjugacy class contains full regular conjugacy 
classes \cite{stan} of all its elements relative the adjoint action 
of the subgroup $G^{\om} \subset G$,
\beq
g \in \cC^{\om}(t) \Longrightarrow G^{\om} g (G^{\om})^{-1} \equiv 
G^{\om} g \om ((G^{\om})^{-1}) \subset \cC^{\om}(t).
\eeq

The remaining part of the original bulk symmetry, $G_L \x G_R$, 
translates - just as in the untwisted case - into covariance
of the ensuing physical model under rigid one-sided translations 
of twisted conjugacy classes within $G$,
\beq
G_L \cC^{\om}(t) G_R = G_L \cC^{\om}(t) \om (G_L^{-1}) \om (G_L) 
G_R = \cC^{\om}(t) G. 
\eeq
This reflects the residual freedom in the definition of the 
boundary state consisting in the choice of the inner \aut twisting 
the gluing condition \cite{auts}. 

Upon specialising the above presentation to the case of $SU(3)$ 
for the sake of illustration and preparation for Sec.\ref{sub:emb}, 
we obtain a classification of twisted branes in terms of twisted 
conjugacy classes in $SU(3)$. For the specific choice 
of the group-integrated \rep of $\om$ given by complex conjugation
$\rho$, we can parametrise the latter as 
\beq\label{tcc-su3-rho}
\cC^{\rho}(\th) = \left\{ h  \lb \barr{ccc}
\cos \th & \sin \th & 0 \\
- \sin \th & \cos \th & 0 \\
0 & 0 & 1 \earr \rb h^T \ \bigg\vert \ h \in SU(3) \ \land \ \th 
\in \left[ 0,\frac{\pi}{2} \right] \right\},
\eeq
from which it transpires that there are two species of twisted 
branes in this background: a $5$-dimensional twisted conjugacy 
class of the group unit, with a maximal stabiliser,\linebreak
$\cC^{\rho}(0) \cong SU(3)/SO(3)$, and generic $7$-dimensional 
twisted conjugacy classes which can be regarded as homogeneous 
spaces $SU(3)/SO(2)$. We shall make an explicit use 
of the parametrisation \refeq{tcc-su3-rho} in the sequel.

\section{Twisted Reflection Equations.}\label{sec:elem}

In this section, we shall discuss (quantum) algebras relevant 
to the description of twisted branes. The arguments we invoke are 
of the kind presented in \cite{qmod}, i.e. they are based on 
the pattern of symmetry breaking induced by twisted branes (cp 
the discussion of the previous section).

Thus we propose to consider a twisted Reflection Equation (tRE):
\beq
\textrm{tRE}^- \quad : \quad \R_{12} \Km_1 \R_{12}^{t_1} \Km_2 = 
\Km_2 \R_{12}^{t_1} \Km_1 \R_{12},\label{tRE1} 
\eeq
in which $\R$ is a bi-fundamental realisation of the standard 
universal $\cR$-matrix  of the relevant quantum group $\qA{2n}$ 
and $\Km$ are operator-valued matrices of generators of the 
twisted Reflection Equation Algebra $\treA{2n}$ (see: the Appendix).

Equations of this kind (parametrised by additional physical 
quantities) have long been known to describe couplings of bulk 
modes to the boundary in $(1+1)$-dimensional integrable models 
on a half-line, with involutively twisted gluing condition 
for chiral symmetry currents at the boundary (see 
\cite{mackay,delius}, and the references within). Furthermore, 
the respective algebraic structures ensuing from \refeq{tRE1} and 
its dynamical counterpart from the papers cited share many essential 
features (coideal property, an intimate relation to the so-called 
symmetric pairs).

The twisted left-right (co)symmetries \cite{qmod} of the tRE: 
$\Km \mapsto \t^T \Km \s$, realised in terms 
of $(\t,\s) \in \GcG \equiv SU_q(2n+1) \ox_{\cR} SU_q(2n+1)$ 
(we have $q = e^{\pi i/(\k + 2n +1)}$, as indicated 
by the underlying CFT), provide a quantum version of the classical 
left-right isometry of the group manifold, which should be 
a symmetry of the problem (to be broken by branes). There is 
another tRE with the same symmetry properties,
\beq
\textrm{tRE}^+ \quad : \quad \R_{21} \Kp_1 \R_{21}^{t_1} \Kp_2 = 
\Kp_2 \R_{21}^{t_1} \Kp_1 \R_{21}\label{tRE2}.
\eeq 
The transformation rule for $\Kp$ reads 
$\Kp\mapsto (S \t) \Kp (S \s)^T$ ($S$ is the antipode of the Hopf 
algebra $SU_q(2n+1)$). As we shall discuss in 
App.\ref{app:qUointREA} and following \cite{nosu}, the two tRE's 
define the same quantum algebra $\qUo{2n+1}$ \cite{tworep}, 
a quantum deformation of $\gt{so}_{2n+1}$. $\textrm{tRE}^\pm$ 
differ in the manner the algebra $\qUo{2n+1}$ is embedded in them. 
In view of the prominent r\^ole played by $SO(2n+1)$ 
in the description of twisted $A_{2n}$ branes (see 
Sec.\ref{sec:tw-Db}.), the appearance of the latter algebra should 
be regarded as an encouraging fact.  

As it turns out \cite{nuw}, we need both $\Kp$ \emph{and} $\Km$ 
to construct Casimir operators for this 
algebra\footnote{In the case at hand, i.e. for the deformation 
parameter $q$ a root of unity, there are - as usual - additional 
central elements in the algebra, originally discovered 
in \cite{haklipo}. They shall not be considered in this paper. 
In particular, for $A_2$ with our subsequent choice of the \rep 
theory, they are known to carry no interesting information 
\cite{klimyk}.}. They shall play an important part in our 
discussion of brane geometries (see Sec.\ref{sub:emb}.). 
The Casimir operators can be cast in the form: 
\beq\label{Kazie}
c_m := \tr \lb \X \lb \bD \X \rb^{m-1} \rb, \qquad m \in 
\ovl{1,2n-1},
\eeq
where $\X := \Km \Kp$ 
and $\bD := \diag (q^{-2 \cdot 2n},q^{-2 \cdot (2n-1)},\ldots,1)$, 
the latter being straightforwardly related to the antipode $S$ 
through 
\beq\label{sqrS}
\bD^{-1} \s \bD = S^2 \s.
\eeq

In the spirit of the papers \cite{qmod,qorb}, we would like 
to identify branes with appropriately chosen \irreps of the tREA 
defined above. Further evidence 
in favour of such an assignment as well as the details 
of the identification shall be provided in Sect.\ref{sec:tgeom}. 
For the present, though, we focus on a particular consequence 
of this idea: clearly, it should entail the existence 
of an algebraic counterpart of \refeq{adj-action}. And indeed, 
the vector part of the $\GcG$ symmetry, realised as
\beq\label{tvecsym2}
\Km \mapsto \s^T \Km \s \qquad ,\qquad \Kp \mapsto (S \s) \Kp 
(S \s)^T  
\eeq
possesses the required properties. In addition to preserving 
the respective tRE's, it also leaves the values of all $c_m$'s 
unchanged. This follows from the fact that under the above 
transformations $\X \mapsto \s^T \X (S \s)^T$, $\X \bD \X \to \s^T \X
(S \s)^T \bD \s^T \X (S \s)^T$ etc. Upon applying \refeq{sqrS}, we
then obtain $\bD^{-1} (S \s)^T \bD \s^T =  \1$ and so we readily  
verify $c_m \mapsto \tr \lb \s^T \X \lb \bD \X \rb^{m-1} (S \s)^T
\rb$. That leads us directly to the conclusion.

Next, we turn to the \rep theory of \refeq{tRE1}-\refeq{tRE2}. 
Recall that $\treA{2n}$ is related to a particular deformation 
of $\gt{so}_{2n+1}$ denoted by $\qUo{2n+1}$.  The \rep theory 
of $\qUo{2n+1}$ is known in considerable detail (see, e.g., 
\cite{tworep,hapos}.). Here, we are interested only 
in the \hw \irrepsn. For $q=e^{\pi i/(\k+2n+1)}$, these are 
of the classical type, with the corresponding highest weights 
truncated to a fundamental domain in a $(\k+2n+1)$-dependent 
way outlined below. We adopt labelling by 
signatures\footnote{The signatures can readily be expressed 
in terms of  the Dynkin labels of the corresponding weights: 
\beq
2m_i = 2 \sum_{j = i}^{n-1} \la_i + \la_n, \quad (i<n), \qquad 
2 m_n = \la_n.\label{mthrul} 
\eeq}: 
$\vec{m} = (m_1,m_2,\ldots,m_n) =: \sum_{i=1}^n m_i \vec{e}_i$ such
that all $m_{i}$'s are integers or all are half-integers, subject to
the dominance condition: 
\beq\label{dom}
m_{1} \geq m_{2} \geq \ldots \geq m_{n} \geq 0.  
\eeq
The truncation scheme has not been worked out in all generality 
as of this writing. It is known \cite{hapos} in the simplest 
case of $\qUo{3}$,
\beq\label{qtrunc}
2 m_{1} \leq \k + 2,
\eeq 
and inspection of the algebra $\qUo{5}$ and its \reps (cp 
\cite{tworep}) reveals that the candidate formula 
is\footnote{At the threshold, matrix elements of the generators 
of $\qUo{5}$ develop poles. Analogous pathology occurs 
for $\qUo{3}$ and extends to Clebsch--Gordan coefficients, as 
well as the associated $6j$-symbols.} $m_{1} + m_{2} \leq \k+5$. 
Thus, it seems plausible that in the general case of \irreps 
of $\qUo{2n+1}$ highest weights are truncated as:
\beq
m_1 + m_n \leq \k+2n+1. 
\eeq
We shall return to this issue in the next section.

\section{Geometry of twisted branes from the 
tREA.}\label{sec:tgeom}

In the present section, we unravel a number of features 
of the tREA's introduced, indicating towards an intimate 
relationship between the latter and twisted branes 
of the WZW models of type $A_{2n}$.

\subsection{Algebraic truncation of twisted brane 
labels.}\label{sub:cutoff}

Below, we address the issue of microscopic localisation 
of twisted branes from two vantage points: the BCFT one, based 
on the notion of a (twisted) boundary state, and that 
of a suitably truncated \rep theory of the $\om_c$-invariant 
subalgebra $\gt{so}_{2n+1}$ which we consistently embed 
in the \rep theory of $\treA{2n}$. The identifications made 
shall then be tested in a semi-classical approximation 
in Sect.\ref{sub:emb}.  

Let us start by recalling that the non-classical geometry 
of a maximally symmetric WZW brane has been successfully  
encoded in the \rep theory of $REA_q(\ggt)$ \cite{qmod,qorb}.
A crucial r\^ole in this approach has been played by the map 
$REA_q(\ggt) \to U_q(\ggt)$ given by $\M = \Lp \M_{0} S \Lm$, 
in which $\mathbf{L}^\pm$ are the familiar FRT operators of 
\cite{frt} (see: the Appendix). The map provides us 
with tools necessary to show that there is a one-to-one 
correspondence between \hw \irreps of $REA_q(\ggt)$ and 
(untwisted) branes. Moreover, it gives geometrical
information about branes in terms of Casimir operators. 

For the tRE, there is a similar embedding 
of $\treA{2n} \cong \qUo{2n+1}$ in $\qA{2n}$,
\beq\label{embtREA}
\Km = \lb \Lp \rb^T \C^- \Lm \quad , \quad \Kp = S \Lp \C^+ 
\lb S \Lm \rb^T, 
\eeq
with $\C$ - a constant ($c$-number-valued) matrix solution 
of tRE. In what follows, we take 
$\C := \diag(c_1,c_2,\ldots,c_{2n+1})$ such that 
$\lim_{q \to 1} c_i = 1, \ i \in \ovl{1,2n+1}$. This choice 
guarantees that in the classical limit, $q \to 1$, 
\refeq{embtREA} defines the embedding: 
\beq\label{standemb}
\mathcal{U}(\gt{so}_{2n+1}) \ni I_{i+1,i} \longmapsto F_i - 
E_i \in \mathcal{U}(\gt{su}_{2n+1}),
\eeq 
in which $I_{i+1,i}$ denote generators 
of $\mathcal{U}(\gt{so}_{2n+1})$. The map \refeq{embtREA} 
determines a branching of \reps $R_{\la}$ of $\qA{2n}$ 
($\Lp,\Lm \in \qA{2n}$) into those of $\qUo{2n+1}$,
\beq\label{qbranch}
R_{\la} \longrightarrow \bigoplus_{\vec m} 
\tilde{b}_{\la}^{\vec{m}} R_{\vec{m}}.
\eeq  
Analogously, the map \refeq{standemb} determines the classical 
counterpart of \refeq{qbranch}. Motivated by the analysis 
of the untwisted case, as well as by the considerations 
of \cite{twfuzzy} and \cite{stan} we propose the following 
identification:\newline\newline
\indent {\bf Twisted branes correspond to those \hw \irreps 
of $\qUo{2n+1}$ which show up on \rhs of \refeq{qbranch}, 
with the branching coefficient $\tilde{b}_{\la}^{\vec{m}}$ 
determining the intersection of the untwisted brane described 
by $R_{\la}$ with the twisted one associated 
to $R_{\vec{m}}$}.\newline\newline
The rule has to be supplemented by a truncation of ${\vec m}$'s 
(denoted by a tilde in \refeq{qbranch}), \emph{stricter} than 
the one on the \hw \irreps of $\qUo{2n+1}$. The truncation is 
imposed on $R_{\vec{m}}$ as detailed below. 
Apart from the truncation, the branching follows the purely 
classical ($q=1$) pattern. It appears that for $\k \in 2\bN^*$ 
one can find a relatively easy algebraic prescription 
for the truncation\footnote{The significance of the parity 
of $\k$ was emphasised already in \cite{stan}.} by demanding 
not only that the number of surviving \irreps agree 
with the number of admissible boundary states from the lattice 
of dominant fractional symmetric affine weights of $A_{2n}$ (cp 
\cite{twcond}), but also that the ensuing distribution 
of $\qUo{2n+1}$-\reps over $\faffA{2n}$ possess the $\bZ_{2n+1}$ 
simple current symmetry of twisted conjugacy classes. It reads  
\beq\label{qcut}
2 m_1 \leq \k
\eeq
and is to be iteratively imposed on the \rep theory 
of $\qUo{2n+1}$ which comes with a tensor product structure 
elucidated in \cite{tensor}. 

Here is a description of the procedure leading to \refeq{qbranch}. 
As the input we use the known \cite{tensor} fact: 
$\tilde{b}_{\La_1}^{\vec{m}} = \d_{\vec m}^{\vec{e}_1}$ 
($R_{\La_1}$ and $R_{\vec{e}_1}$ are the fundamental \reps 
of $\qA{2n}$ and $\qUo{2n+1}$, respectively). The procedure is 
itarative. Let 
$R_\la = \bigoplus_{\vec m} \tilde{b}_{\la}^{\vec{m}} R_{\vec{m}}$ 
be known (we start with $R_{\La_1}$). In a single step, we 
tensor $R_\la$ with $R_{\La_1}$. On the $\qA{2n}$ side, 
this yields 
$R_\la \ox R_{\La_1} = \bigoplus_{\mu \in \faffA{2n}} 
\cN_{\la,\La_1}^{\ \ \mu} R_\mu$ ($\cN_{\la,\La_1}^{\ \ \mu}$ are 
multiplicities). On the $\qUo{2n+1}$ side, we get 
$\bigoplus_{\vec m} \tilde{b}_{\mu}^{\vec{m}} R_{\vec{m}} \ox R_{\vec{e}_1}$. 
Luckily  \cite{tensor}, tensor products of the kind 
$R_{\vec{m}} \ox R_{\vec{e}_1}$ are 
well-defined\footnote{Due to the fact that $\qUo{2n+1}$ is 
a coideal (non-Hopf) subalgebra of $\qA{2n}$ tensoring is 
problematic in general.} and can be decomposed into irreducible 
components. We may then derive the branching coefficients 
$\tilde{b}_\mu^{\vec{m}}$ for the irreducible simple summands 
$R_\mu$ upon imposing the truncation \refeq{qcut}. Clearly, we 
can reconstruct the entire \rep theory of $\qA{2n}$ 
over $\faffA{2n}$ in this way, hence we retrieve all the desired 
intersections. 

Several comments are well due at this point. First of all, our 
usage of the quantum algebras should not obscure the fact that 
the truncation could just as well be imposed in the classical 
setup (i.e. for $\gt{so}_{2n+1}$). The good news is that it can 
be reconciled with the specific structure of the \rep theory 
of $\qUo{2n+1}$ for $q$ a root of unity. Indeed, in consequence 
of \refeq{dom}, the present truncation $2 m_1 \leq \k$ implies 
$m_1 + m_n \leq 2 m_1 \leq \k < \k + 2n + 1$ and hence it is more 
restrictive than \refeq{qtrunc}. Finally, the \reps admitted 
by \refeq{qbranch} correspond to those \reps of the algebra 
$\gt{so}_{2n+1}$ which can be integrated to \reps of the group 
$SO(2n+1)$ \cite{twfuzzy}. The latter fact shall be of prime 
relevance to the discussion of the next section.

Let us also note another, rather astonishingly exact correspondence 
between \refeq{qbranch} and BCFT. Namely, we can 
calculate\footnote{Details of the relevant BCFT computation leading 
to \refeq{scovl} shall be presented in an upcoming paper.} scalar 
products of a twisted boundary state $\ketCo{\dot \mu}$ with all 
admissible untwisted boundary states $\ketC{\la}$, whereby we obtain
\beq\label{scovl}
\braCo{\dot \mu} \ketC{\la} = \lb n_{\la}^{\om_c} \rb_{\Psi(0)}^{\dot
  \mu}, \qquad \Psi(0) := \frac{1}{2} E \lb \frac{\k}{2} \rb \lb \La_n
+ \La_{n+1} \rb.  
\eeq
Here, $n_{\la}^{\om_c}$ are the so-called twisted fusion rules 
of the CFT \cite{twfus} and $E(x)$ denotes the integral part 
of $x$. It appears that for even $\k$ the branching coefficients of
\refeq{qbranch} coincide with the twisted fusion rules as 
\beq
\tilde{b}_{\la}^{\vec{m}} = \lb n_{\la}^{\om_c}
\rb_{\Psi(0)}^{\Psi(\vec m)},  
\eeq 
with the identification between the truncated \rep theory 
of $\treA{2n}$ and the set of twisted boundary labels given 
by the mapping:
\beq\label{quelmap0}
\Psi \ : \ \vec m \longmapsto \Psi(\vec m) := \frac{1}{2} 
\sum_{i=1}^{n-1} (m_{n-i} - m_{n-i+1})(\La_i + \La_{2n+1-i}) + 
\frac{\k - 2 m_1}{4}(\La_n + \La_{n+1}), 
\eeq
originally proposed in \cite{quel} and further discussed 
in \cite{twfuzzy}. Thus \refeq{quelmap0} completes our translation 
of the BCFT data into the quantum-algebraic language of the tREA.
Note that it actually associates (through \refeq{qbranch} 
and \refeq{scovl}) the trivial \repn, $R_{\vec 0}$, 
with the dimensionally reduced twisted brane (the one wrapping 
the twisted conjugacy class of the group unit) as the unique one 
having a non-vanishing overlap with (i.e. containing) the pointlike 
untwisted branes localised at the $2n+1$ points in $SU(2n+1)$ 
corresponding to the elements of the centre 
$Z(SU(2n+1)) \cong \bZ_{2n+1}$. We shall come back to this point 
in the next section.

\subsection{Brane localisation from Casimir 
eigenvalues.}\label{sub:emb}

We are not aware of any natural embedding 
$\treA{2n} \emb \reA{2n}$. Recall that - following \cite{qmod} - we 
assign to the latter algebra the r\^ole of the quantised algebra 
of functions on the group manifold. Thus, the lack of such a map 
prevents us from giving a direct geometrical meaning to various 
quantities associated with tREA's, e.g. to their Casimir operators. 
Luckily, the situation is not hopeless. We may employ \refeq{embtREA} 
and the map $\reA{2n} \to \qA{2n}$, \refeq{embREA}, to construct 
a map $\treA{2n} \to \reA{2n}$ order by order in the parameter $1/\k$, 
in a manner consistent with the $q \to 1$ limiting procedure described 
in \cite{qmod}. Using the above expansion we shall express 
the quadratic Casimir operator $c_1$ of $\treA{2n}$ in terms 
of the $\M$-variables, that is in terms of solutions 
to the (untwisted) RE (cp \cite{qmod,qorb}). All approximate 
equalities below are up to terms of higher order 
in the expansion parameter. We also choose $\C:=\1$.                     

First, note that $K^\pm_{ii} \approx \1$ for all 
$i \in \ovl{1,2n+1}$. Hence 
$c_1 \approx \sum_i \1 + \sum_{i>j} K^-_{ij} K^+_{ji}$. 
Upon subtracting the trivial part, we then define
\def\ct{{\tilde c}}
\beq
\ct_1 := \sum_{i>j=1}^{2n+1} K^-_{ij} K^+_{ji}.
\eeq
We also have 
$K^-_{ij} \approx \sum_{j\leq k\leq i} L^+_{ki} L^-_{kj}$ 
and 
$K^+_{ji} \approx \sum_{j\leq k\leq i} SL^+_{ik} SL^-_{jk}$. 
Using the results from App.D of \cite{qorb} we list 
the relevant (leading) terms of the $\mathbf{L}^\pm$-operators:
\beq\barr{lcll}
L^+_{ij} \approx \la E_{ji}& \quad ,& \quad SL^+_{ij} 
\approx -\la E_{ji},& \quad i<j \\ \non
L^-_{ij} \approx -\la E_{ji}& \quad ,& \quad SL^-_{ij} 
\approx \la E_{ji},& \quad j<i \\ \non
L^\pm_{ii} \approx \1 & \quad ,& \quad SL^\pm_{ii} \approx 
\1,&
\earr
\eeq
with $E_{ij}$ defined as in \cite{qorb} (their explicit form 
is not relevant here). The above yield
\beq
K^-_{ij} \approx \la (E_{ij}-E_{ji}) \approx - K^+_{ji}, 
\qquad j<i
\eeq
and - since $M_{ij} \approx \la E_{ji}$ for $i\neq j$ - we 
conclude that
\beq
K^{\mp}_{ij} \approx M_{ij} - M_{ji}. 
\eeq
Thus
\beq\label{ccm}
\ct_1 \approx -\sum_{i>j=1}^{2n+1}(M_{ij}-M_{ji})^2=\frac12 
\tr(\M-\M^T)^2.
\eeq
At this stage, we may already evaluate the Casimir operator on 
a particular \irrep $R_{\vec{m}}$ of $\treA{2n}$. Thus we 
rewrite \lhs after \cite{tworep,CasoN} in terms of components 
of the signature vector $\vec{m}$ labelling the \irrep 
chosen, whereby we obtain
\beq\label{reCas2m}
\ct_1\big\vert_{R_{\vec{m}}} = q^{2n-1} \la^2 
\sum_{j=1}^n [m_{n+1-j}]_q[m_{n+1-j}+2j-1]_q.
\eeq 
On the present level of generality, we may draw one encouraging 
conclusion: the Casimir operator clearly vanishes on the trivial 
\rep of the tREA, $R_{\vec 0}$, and with our choice of truncation 
of admissible \irrepsn, \refeq{qcut}, it is also 
the unique\footnote{Note that \refeq{qtrunc} does not guarantee 
the uniqueness.} \rep with this property. Thus for $\vec m = \vec 0$
we obtain: $\tr(\M-\M^T)^2 \propto \ct_1 = 0$, which is solved by
symmetric matrices $\M$. This conforms with the known results for the
dimensionally reduced brane \cite{mms} to which we consequently
associate the zero $\qUo{2n+1}$-signature, consistently with our
microscopic analysis. Equivalently, from the (co)isometry
\refeq{tvecsym2} of \irreps of $\qUo{2n+1}$ we conclude that the
geometry defined by $R_{\vec 0}$ is encoded in the twisted
$SU_q(2n+1)$-comodule algebra: $\C \mapsto \s^T \C \s$ and therefore
it describes the twisted (quantum) conjugacy class of the group unit.

It turns out that we may extract further information 
from the semiclassical result \refeq{ccm}-\refeq{reCas2m}, whereby 
we gain some insight into its physical meaning. To these ends we 
specialise the formul\ae ~to the simplest physically 
relevant\footnote{The classical $SU(2)$ has no non-trivial diagram 
\autsn.} case: $n=1$. Plugging into \refeq{ccm} the explicit 
classical parametrisation \refeq{tcc-su3-rho} of twisted conjugacy 
classes of $G = SU(3)$,
\beq\label{mat} 
M_{\th} =  \lb \barr{ccc}
\cos \th & \sin \th & 0 \\
- \sin \th & \cos \th & 0 \\
0 & 0 & 1 \earr \rb, 
\eeq 
and comparing with \refeq{reCas2m} we get the relation:
\beq\label{Cas2for3} 
 - 8 \sin^2 \th = 2 \la^2 [\la_1/2]_q [\la_1/2+1]_q, 
\eeq 
where - as previously - $\la_1 = 2m_1 \in \bN$ \cite{tworep}. 
We can regard \refeq{Cas2for3} as a quantisation condition for 
brane positions. For $1 \ll \la_1 \ll \k$ it yields 
\beq\label{thql} 
\th \approx \frac{\la_1 \pi}{2\k}.  
\eeq 
Clearly, the above rule retains its validity for $\la_1 = 0$, 
hence we may expect it to be generally applicable in the large 
$\k$ limit. 

The significance of the classical limit \refeq{thql} of our 
quantum-algebraic result follows from the fact that it is amenable 
to direct comparison with the data on twisted brane localisation 
which can be found in the literature\footnote{As for exact BCFT 
data of, e.g., \cite{mms} it unavoidably becomes obscured 
by the conventions adopted in the original papers. They differ 
from ours in the choice of the representative of the class of \auts 
implementing the Dynkin diagram reflection on the group level.}. 
Thus we compare \refeq{thql} with the relative-cohomological 
analysis of \cite{stan}, using the same group-integrated
representative of $\om_c$ as the one quantised by the tRE's
\refeq{tRE1}-\refeq{tRE2}. The analysis yields a quantisation rule: 
\beq
\th = \frac{(2 n - \k)\pi}{2 \k}, \quad n \in 
\ovl{E\lb\frac{\k}{2}\rb,\k},
\eeq
which falls in perfect agreement with \refeq{thql} (for even $\k$) 
and, consequently, lends support to our proposal. Indeed, 
upon restricting in \refeq{reCas2m} to integer-spin \irreps 
of $\qUo{3}$, the two quantisation formul\ae ~become fully 
equivalent. The latter \repsn, on the other hand, are precisely 
the ones that appear in (truncated) branchings of the \irreps 
of $\reA{2}$ used in \cite{qorb} in the description of untwisted 
branes, as determined by \refeq{qbranch}.

\section{Summary and conclusions.} 

In the present paper, we have discussed a class of quantum 
algebras, the twisted Reflection Equation Algebras 
$\treA{2n}$, in reference to twisted boundary states 
of WZW models for the groups $SU(2n+1)$ and the associated 
brane worldvolumes wrapping (classically) twisted conjugacy 
classes within the group manifolds. The framework, developed 
as a straightforward extension of the previous constructions 
for untwisted WZW branes, based on the untwisted Reflection 
Equation Algebras $\reA{2n}$, is a novel proposal for a compact 
algebraic description of the twisted branes. Our study provides 
several arguments in favour of its profound relationship 
to the CFT of twisted boundary states: classical-type \irreps 
of $\treA{2n}$ enjoy a (co)symmetry that quantises the twisted 
adjoint symmetry of the boundary states (the starting point 
of the construction) and in so doing they realise a symmetry 
breaking scenario analogous to the BCFT one (cp the introductory 
remarks under \refeq{glue0} and \refeq{adj-action}); 
the eigenvalues of the Casimir operators of $\treA{2n}$ returned 
by these \irreps admit a simple physical interpretation in terms 
of quantum localisation rules for twisted brane geometries, shown 
to reproduce the known result for the simplest case of $SU(3)$ 
in the semiclassical approximation allowing for an explicit 
embedding $\treA{2n} \emb \reA{2n}$; the \rep theory 
of $\treA{2n}$, endowed with a restricted tensor product structure 
remarked upon under \refeq{qcut}, seems to reproduce microscopic 
twisted brane density distributions within the quantum manifolds 
of the $SU(2n+1)$ upon truncating the set of admissible dominant 
signatures (labels of the \irrep of $\treA{2n}$); the truncation 
is identical with the one suggested in \cite{quel} in the BCFT 
context. 

In conclusion, we believe that there are sound reasons to regard 
the tREA's as natural building blocks of quantum-algebraic 
matrix models for twisted branes on the $SU(2n+1)$ WZW 
manifolds. While encouraged by the results obtained hitherto, 
we are aware of numerous questions that our study leaves 
unanswered, such as the harmonic analysis on the associated 
geometries, and the gauge dynamics of twisted WZW branes that 
the algebras are claimed to describe. We intend to return to them 
in a future publication.

\acknowledgments  
The authors would like to thank the organisers of the 2004 ESI 
Workshop on "String theory on non-compact and time-dependent 
backgrounds", where part of this work was done. R.R.S. gratefully 
acknowledges useful discussions with Thomas Quella and essential 
help with the numerics involved from Marta A.~Hallay--Suszek. 
J.P. expresses his gratitude to the String Theory Group at Queen 
Mary London, and especially to Sanjaye Ramgoolam and Steven Thomas, 
for their kind hospitality during J.P.'s visit. It is also a
pleasure to thank the Theoretical Physics Group at King's College
London, and in particular Andreas Recknagel, Sylvain Ribault and
Thomas Quella, for their interest in the project, discussions and a
stimulating atmosphere in the final stage of this work.\newpage

\appendix

\section{(Twisted) Reflection Equations.}\label{app:tRE}

In this appendix, we discuss chosen properties of three RE's:
\beqa
\textrm{tRE}^- \quad : \quad \R_{12} \Km_1 \R_{12}^{t_1} \Km_2 &=& \Km_2 
\R_{12}^{t_1} \Km_1 \R_{12},  \label{tRE1app} \\ \non
\textrm{tRE}^+ \quad : \quad \R_{21} \Kp_1 \R_{21}^{t_1} \Kp_2 &=& \Kp_2 
\R_{21}^{t_1} \Kp_1 \R_{21}, \label{tRE2app} \\ \non
\textrm{RE0} \quad : \quad \R_{12} \M_1 \R_{21} \M_2 &=& \M_2 \R_{12} \M_1
\R_{21},  
\label{RE0}
\eeqa
appearing in the paper. In the formul\ae ~above, $\R$ is a bi-fundamental realisation  
of the standard universal $\cR$-matrix  of the relevant quantum group $\qA{2n}$,
$\R \equiv \lb R_V \ox R_V \rb ( \cR )$,
satisfying the celebrated Quantum Yang--Baxter Equation (see, e.g., \cite{press}).
The operator-valued matrix $\K^{\mp}$ (resp. $\M$) generates the twisted (resp. 
untwisted) Reflection Equation Algebras, $\treA{2n}^{\mp}$ (resp. $\reA{2n}$) whose quantum 
group comodule structure and relation to twisted (resp. untwisted) quantum algebra 
$\qUo{2n+1}$ (resp. $\qA{2n}$) shall be discussed in the sequel.

\subsection{Symmetries of the RE's and their relation to $\qA{2n}$.}  

The three RE's of interest enjoy the following (twisted) left-right
(co)symmetries  
which are crucial for their applicability in an effective
description of  
branes in WZW models ($S$ is the antipode of the Hopf algebra $SU_q(2n+1)$):
\beqa\label{lrsym2} 
&\Km \mapsto \t^T \Km \s \quad , \quad \Kp \mapsto (S \t) \Kp (S \s)^T,
\\ \non
&\M \mapsto \t \M S \s,
\eeqa
where 
\beq \R_{12} \s_1 \s_2 = \s_2 \s_1 \R_{12},\quad
\R_{12} \t_1 \t_2 = \t_2 \t_1 \R_{12},\quad \R_{12} \t_1 \s_2 = \s_2
\t_1 \R_{12} \label{st3} 
\eeq
are the defining relations of (two copies of) the quantum group $SU_q(2n+1)$ 
associated to the $\cR$-matrix $\R$.
 
Solutions to the three RE's under study can straightforwardly be realised in
terms of generators of the (extended) quantum universal enveloping algebra 
$\qA{2n}$ through 
\beqa 
&\Km = \lb \Lp \rb^T \C^- \Lm \quad , \quad \Kp = 
\lb S \Lp \rb \C^+ \lb S \Lm \rb^T, \label{embtREAapp} \\ \non 
&\M = \Lp \M_{0} S \Lm, \label{embREA}
\eeqa
where $\C^{\mp}$ and $\M_0$ denote respective (arbitrary) constant solutions 
($c$-number-valued matrices) and $\Lpm$ are the familiar FRT-operators
\cite{frt}. The existence of the homomorphisms thus defined enables us 
to use the well-known \rep  theory of the quantum algebra $\qA{2n}$ to induce 
a \rep theory of the (t)REA's. In particular, the relevant (specialised) \rep 
theory of $\qA{2n}$ has been studied at some length in \cite{qorb}.

\subsection{The two embeddings $\qUo{2n+1} \emb \treA{2n}^{\mp}$.}
\label{app:qUointREA} 

The twisted quantum orthogonal algebra $\qUo{2n+1}$, considered originally by 
Gavrilik and Klimyk in \cite{tongk}, is defined by the following commutation
relations:
\beqa
&[\Pi_i,\Pi_j] = 0 \quad \textrm{if} \quad \vert i - j \vert > 1, \\ \non
&\Pi_i^2 \Pi_j - \2 \Pi_i \Pi_j \Pi_i + \Pi_j \Pi_i^2 = - \Pi_j \quad \textrm{if} 
\quad \vert i - j \vert = 1,
\eeqa
satisfied by its generators $\Pi_i, \ i \in \ovl{1,2n+1}$. In the classical
limit,  
$q \to 1$, the above relations reproduce the standard defining relations of 
$\Uo{2n+1}$. They differ, on the other hand, from the defining relations of the 
quantum universal enveloping algebra $\cU_q (\gt{so}_{2n+1})$ (of Drinfel'd and 
Jimbo) associated to the universal $\cR$-matrix for $\gt{so}_{2n+1}$ (e.g.\cite{press}).

In addition to the above generators, we define after \cite{nuw} the operators 
$\Pi^{\mp}_{ji}, \ 1 \leq i < j \leq 2n+1$ through:
\beqa
&\Pi^{\mp}_{i+1,i} := \Pi_i, \non \\ \nonumber
&\Pi^{\mp}_{ji} := \Pi^{\mp}_{jk} \Pi^{\mp}_{ki} - q^{\mp 1} \Pi^{\mp}_{ki} 
\Pi^{\mp}_{jk} \quad \textrm{for arbitrary} \quad i < k < j.
\eeqa
It is then a matter of straightforward algebra to verify that the elements of the 
two operator-valued solutions to \refeq{tRE1app}-\refeq{tRE2app} provide a
realisation  
of the algebra of $\Pi^{\mp}_{ji}$'s. More precisely, we have the identification:
\beq\label{sontrea}
K_{ij}^- = \la q^{2n-j} \Pi^-_{ij} \quad , \quad K_{ij}^+ = - \la q^{2n+1-j} 
\Pi^+_{ji},
\eeq
establishing a homomorphism $\qUo{2n+1} \emb \treA{2n}^{\mp}$. This, together
with the explicit mappings $\treA{2n}^{\mp} \to \qA{2n}$, \refeq{embtREAapp}, 
embeds $\qUo{2n+1}$ in $\qA{2n}$ as the so-called coideal subalgebra \cite{nosu}. 
Its \rep theory, both of classical and non-classical type, has been discussed 
in great detail in a series of papers \cite{tongk,tworep,hapos,tensor}, also 
in relation to the \rep theory of $\qA{2n}$. An important conclusion following 
from that analysis is that we can effectively restrict to $\qUo{2n+1}$-\irreps 
of the classical type as long as we are dealing with classical-type \irreps 
of $\qA{2n}$ (branching into the former).

\end{document}